\documentclass{article}
\usepackage{ismir,amsmath,cite,url}
\usepackage{graphicx}
\usepackage{color}
\usepackage{booktabs}

\usepackage{amsmath,amssymb,amsfonts,amsthm}
\usepackage{subcaption}
\usepackage{color}
\usepackage{caption}
\usepackage{multirow} 
\usepackage{arydshln} 
\usepackage{todonotes}
\usepackage{pgf}

\usepackage[explicit]{titlesec}
\title{Deep Unsupervised Drum Transcription}
\newcommand{\largesqueeze}{\vspace{-0.2cm}}
\newcommand{\squeeze}{\vspace{-0.15cm}}
\newcommand{\smallsqueeze}{\vspace{-0.1cm}}

\twoauthors
{Keunwoo Choi} {Spotify \\ {\tt keunwooc@spotify.com}}
{Kyunghyun Cho} {New York University, Facebook AI Research \\
{\tt kyunghyun.cho@nyu.edu}}

\begin{document}
\maketitle
\squeeze

\begin{abstract}
\squeeze
We introduce DrummerNet, a drum transcription system that is trained in an unsupervised manner. DrummerNet does not require any ground-truth transcription and, with the data-scalability of deep neural networks, learns from a large unlabeled dataset. In DrummerNet, the target drum signal is first passed to a (trainable) transcriber, then reconstructed in a (fixed) synthesizer according to the transcription estimate. By training the system to minimize the distance between the input and the output audio signals, the transcriber learns to transcribe without ground truth transcription. Our experiment shows that DrummerNet performs favorably compared to many other recent drum transcription systems, both supervised and unsupervised.
\end{abstract}

\squeeze
\section{Introduction}
\label{sec:introduction}
\smallsqueeze
Transcription is a music information retrieval task with the goal of estimating the score $y$ when input audio $x$ is given. The majority of the recent transcription systems is based on supervised learning, where the transcriber is an \textit{analysis} system 
$\hat{y} = F_a(x)$ that is trained with annotated pairs $\{ (x_m, y_m) \}_{m=1}^M$ to minimize the distance between $y$ and $\hat{y}$ \cite{poliner2006discriminative, bock2012polyphonic, sigtia2016end, vogl2016recurrent, vogl2017drum, southall2016automatic, southall2017automatic, cartwright2018increasing}. 

The trend is similar in drum transcription on which we focus in this paper. 
Supervised learning approaches may incorporate models based on frame-based feature extraction and classification \cite{gouyon2000use}, non-negative matrix factorization (NMF) for pattern matching
\cite{dittmar2014real}, or hidden-Markov model
~\cite{paulus2009drum}. 
More attention has been given recently to deep learning based models such as convolutional neural networks (CNNs, \cite{gajhede2016convolutional, southall2017automatic}) and recurrent neural networks (RNNs, \cite{vogl2016recurrent, vogl2017drum, southall2016automatic}), all of which have greatly improved transcription systems.

However, the lack of a large-scale annotated dataset is one of the most frequently mentioned obstacles that hinder further improvement. 
In practice, this limits the generalizeability of supervised learning systems, as will be discussed in Section~\ref{sec:experiment},
and using synthetic data is one way to address this issue \cite{cartwright2018increasing, vogl2018towards}. Although there have been proposals to use unlabeled data \cite{wu2017automatic, wu2018labeled}, the issue remains as they still rely on supervised learning combined with teacher-student learning \cite{hinton2015distilling}. 
Parallel to those approaches, an annotation-free and, therefore, a more scalable and generalizable alternative would be unsupervised learning. 

Unsurprisingly, one of the humans' music learning procedures, self-taught by trial-and-error, is very similar to unsupervised learning. For example, musicians learn to transcribe by (a) listening, (b) playing an instrument, (c) identifying differences, and (d) making adjustments. \textit{Can this be done without any supervision?} \textit{Yes}, if the person can spot the pitch difference (e.g., the pitch should be higher or lower). Consistent with this logic, developing a transcription system based on unsupervised learning would be feasible if the system can test the estimation, measure the error, and correct itself accordingly.

To implement such an unsupervised transcription system, we need a \textit{synthesis} system, $\hat{x}=F_s(\hat{y})$, making the overall system $\hat{x}=F_s(F_a(x))$. During its training, the system is given $\{x\}_{m=1}^M$ and trained to minimize the distance between $x$ and $\hat{x}$.
There have been few systems relying on unsupervised learning as explained above.
In MIR, the system in \cite{abdallah2006unsupervised} utilized sparse coding to learn a dictionary of time-frequency templates of piano and harpsicord, assuming a (matrix-)multiplication model with additive noise, $F_s(\mathbf{y})=\mathbf{Ay}+\mathbf{e}$. Yoshii~et~al. proposed to use sparse coding in a jointly-learned chord recognition and transcription system ~\cite{yoshii2012unsupervised}. Berg~et~al. designed a probabilistic graphical model that parameterizes the spectral and temporal envelopes, note events, and note activations, in order to transcribe piano by inferring their parameters ~\cite{berg2014unsupervised}.
In drum transcription, many systems have used NMF to decompose a drum track spectrum into spectral templates and their temporal activations (or transcription) \cite{paulus2005drum, wu2015drum}. Several variants of NMF were proposed to address the limits of the fixed spectrum template of NMF \cite{roebel2015automatic, laroche2017drum, lindsay2012drumkit}.
Lastly, a similar system can be found in computer vision, where the parameters of input images are estimated by reconstruction using an image renderer \cite{kanazawa2018learning}.

In this paper, we introduce DrummerNet, a deep neural network based drum transcription system that is trained by unsupervised learning. With a more end-to-end approach, DrummerNet is distinguished from previous research \cite{abdallah2006unsupervised, yoshii2012unsupervised, berg2014unsupervised}, which has strong priors on the target sounds. In \S \ref{sec:systemdesign}, we present the system design principle behind DrummerNet, followed by its details in \S \ref{sec:proposed_method}. In \S \ref{sec:experiment}, the evaluation results are discussed along with the ablation study. We present our conclusion, the problems of our system, and the future direction towards fully unsupervised learning in transcription/MIR in \S \ref{sec:conclusion}.

\squeeze
\section{System Design Principles} \label{sec:systemdesign}
\smallsqueeze

\begin{table}[t]
\centering
\small
\begin{tabular}{lll}
Name                      & Description                  & Note                 \\ \toprule
$n, N$ & The temporal index/length of audio input &  \\ 
$k, K$ & The index/total number of drum components  & $K$$=$$11$ \\ 
$x$, $y$ & Mixture and transcription & $\in \mathbb{R}^N$\\ 
$\hat{x}$, $\hat{y}$ & Estimations of mixture/transcription & $\in \mathbb{R}^N$\\ 

\end{tabular}
\caption{Symbols used in this paper}
\label{my-label}
\end{table}

\begin{figure}[t]
\centering
 \includegraphics[width=\columnwidth]{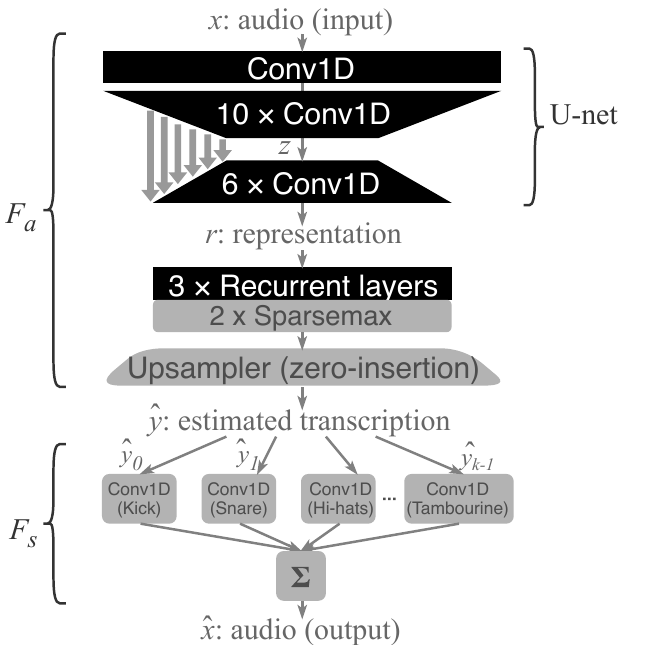}
 \caption{Block diagrams of DrummerNet structure. Trainable modules are illustrated as black boxes and fixed modules as rounded grey boxes.}
 \label{fig:overview}
\end{figure}

Training the proposed DrummerNet is similar to the previous unsupervised learning approaches in music \cite{abdallah2006unsupervised, yoshii2012unsupervised, berg2014unsupervised}, as they all train a system to output $\hat{x}$ that reconstructs the original signal $x$. The difference between $\hat{x}$ and $x$ works as a proxy of the difference between $\hat{y}$ to $y$.

There are three conditions under which unsupervised learning of a transcriber can be achieved successfully.
First, the output of the analysis module $F_a$ must be in the form of transcription -- a set of discrete events representing the timing and intensity of the notes. Second, the synthesis module $F_s$ must synthesize the audio signal given the transcription input $\hat{y}$.
Third, all the components in the network must be differentiable as we rely on backpropagation to train it.

\squeeze
\section{DrummerNet}
\label{sec:proposed_method}
\smallsqueeze

In this section, we introduce the proposed system structure. We specify the number of channels, kernel size, and stride as \texttt{(channel, kernel, stride)}. All the convolutional and recurrent layers use an exponential linear unit as an activation function \cite{clevert2015fast}. \footnote{The implementation of DrummerNet is available on \url{https://github.com/keunwoochoi/DrummerNet}}

\subsection{Analysis module $F_a$}
The analysis module $F_a$, as illustrated in the top half of Figure \ref{fig:overview}, takes the audio signal $x$ as an input and processes it through a series of U-net variant \cite{ronneberger2015u}, recurrent layers, and gated Sparsemax activation \cite{martins2016softmax}. After training, this module is used as a transcriber (with peak-picking).

\largesqueeze
\paragraph{U-net}
The U-net consists of 1D convolutional layers, max-pooling layers, and concatenations between the encoder and the decoder. The encoder consists of a convolutional layer \texttt{(128, 3, 1)} followed by 10~convolutional layers \texttt{(50, 3, 1)} interleaved with max-pooling of size 2. As a result, it outputs $z \in \mathbb{R}^{N/1024}$ which has a receptive field size of 3,072~time steps.

The decoder has only 6~convolutional layers \texttt{(50, 3, 1)} interleaved with a concatenation with the feature map at the same depth as in the encoder and a  $\times$2 bi-linear interpolation. We call the output of decoder $r \in \mathbb{R}^{N/16}$, the representation based on which the transcription is estimated.
The asymmetry between the encoder and the decoder makes the length $r$ to be shorter by a factor of $4^2=16$ compared to that of input $x$. Assuming the input audio is sampled at 16~kHz,\footnote{
This is the sampling rate of input audio in our experiment.
} $r$ would have a sampling rate of 1,000~Hz. 
\largesqueeze
\paragraph{Recurrent layers}

We use three recurrent layers: ($\text{GRUs}$ \cite{chung2014empirical}) \{along time-axis, bi-directional, 100-channel\}, \{along time-axis, uni-directional, 50-channel\}, and \{along channel-axis, uni-directional, 
$K$-channel\}. These three recurrent layers have properties of  \romannumeral 1)~being bi-directional so that the onset at $n$ can be determined by the vicinity of $n$ (both the past and the future), \romannumeral 2)~enforcing temporal dependency, and \romannumeral 3)~enforcing component-wise dependency, respectively. The width (or the hidden vector size) of the third recurrent layer is equal to $K$, the number of drum components in the synthesizer, to map each channel to each drum component.
\largesqueeze
\paragraph{Sparsemax}

In an ideal case of transcription, there would be local sparsity along both the time and channel-axes because the drum events would not repeat with a rate of 1,000~Hz (which is faster than 16-beat on 240~BPM), nor would all the $K$ drum components be activated simultaneously. Although sparsity is one of the properties that \textit{can} be achieved by the autoregressive nature of the recurrent layers, we add Sparsemax \cite{martins2016softmax} activation to encourage it explicitly. The output of Sparsemax has two important properties:  \romannumeral 1) it always sums to 1 (same as Softmax) and \romannumeral 2) it is highly likely to be sparse with actual zeros (unlike Softmax). In DrummerNet, two Sparsemax layers are applied in parallel, one along channel-axis (=instrument-axis) and the other time-axis within a non-overlapping window size of 64. This design choice is based on the assumption that there are only a few onsets among notes (channel-axis sparsity) and within 64~samples at $\hat{y}$, or 64~ms (temporal sparsity). The outputs from these two Sparsemax layers are then multiplied element-wise. 

\largesqueeze
\paragraph{Upsampler} 

Finally, the low temporal resolution of the Sparsemax output is addressed by zero-insertion upsampling by the factor of 16. According to this, we modify the temporal quantization rate of events, unlike the upsampling of a digital signal.

\subsection{Synthesis module $F_s$}
\color{black}

The synthesis module $F_{s}$ consists of $K$ parallel 1D convolutional layers and a channel-wise summing operator. The kernel of each layer is not trained but fixed to the known waveform of each drum component to convert a transcription of a component $\hat{y}_k$ into a track $\hat{x}_k$. The tracks are summed to generate the final output $\hat{x}$ ($=\sum_{k=1}^{K} \hat{x}_k$), the synthesized audio signal. This module is only used during training.

In the implementation, we use $K=11$, using Subclass in Table~\ref{table:components}, following \cite{southall2017mdb}. Ones marked with asterisks were excluded due to their scarcities in our source of isolated drum recordings, which consisted of 12~virtual drum instruments provided by Logic~Pro~X. Multiple drum kits, including rock, pop, funk, and soul\footnote{
Brooklyn, Heavy, Liverpool, Neo~Soul, Detroit Garage, Motown Revisited, Portland, Sunset, Speakeasy, SoCal, Smash, and Slow Jam. All with velocity=98.}, 
were used to prevent the network from overfitting to a specific drum kit. During training, a drum kit was randomly assigned for every batch. 

\begin{table}[t]
\begin{tabular}{lll}
Class & Subclass                                                     & Description                                                                       \\ 
\toprule
KD    & KD                                                           & Kick drum                                                                         \\ 
\hline
SD    & SD                                                           & Snare drum                                                                        \\ 
\hline
HH    & CHH, PHH                                                     & Closed/pedalled hi-hat                                                            \\
      & OHH                                                          & Open hi-hat                                                                       \\ 
\hline
TT    & 
    HIT, MHT, & High/high-mid/ \\
      & HFT,  \color{gray} LFT* \color{black} & high-floor/low-floor tom \\
    
\hline
CY    & RDC, \color{gray}RDB* \color{black}                                                    & Ride cymbal, ride cymbal bell                                                     \\
      & \begin{tabular}[c]{@{}l@{}} \color{black}CRC, \color{gray} CHC*,\\ \color{gray} SPC*\color{black}\end{tabular}      & \begin{tabular}[c]{@{}l@{}}Crash/china cymbal\\ splash cymbal\end{tabular}        \\ 

\hline
OT    &\color{gray} SST*\color{black}, TMB                                                     & \color{black} side stick, tambourine                                                           
\color{black}
\end{tabular}
\caption{A drum component hierarchy \cite{southall2017mdb}. The synthesizer $F_s$ consists of 11 classes, following Subclass of the table with omitting ones marked with asterisks *.}
\label{table:components}
\end{table}
\color{black}

\begin{table}[t]
\begin{tabular}{lll}
Module                          & Input (size)                  & Output (size)                 \\ 
\toprule
U-net encoder                   &                               &                               \\
\hspace{0.3cm}Conv1D            & $x$:$(1, N)$                  & $(C*, N)$                     \\
\hspace{0.3cm}$10 \times$ Conv1D & $(C*, N)$                     & $(C*, N/1024)$                \\ 
\hline
U-net decoder                   &                               &                               \\
\hspace{0.3cm}$6 \times$ Conv1D & $(C*, N/1024)$                & $r$:$(C*, N/16)$              \\ 
\hline
Recurrent layers                & $(C*, N/16)$                  & $(K, N/16)$                  \\ 
\hline
Sparsemax                       & $(K, N/16)$                  & $(K, N/16)$                   \\ 
\hline
Upsampler                       & $(K, N/16)$                   & $\hat{y}$:$(K, N)$            \\ 
\hline
Synthesis module                &                               &                               \\
\hspace{0.3cm}Channel splitter  & $\hat{y}$:$(K, N)$            & $K \times \hat{y}_k$:$(1, N)$ \\
\hspace{0.3cm}Each Conv1D       & $\hat{y}_k$:$(1, N)$          & $\hat{x}_k$:$(1, N)$          \\
\hspace{0.3cm}Sum (mixer)       & $K \times \hat{x}_k$:$(1, N)$ & $\hat{x}$:$(K, N)$           
\end{tabular}
\caption{The shapes of inputs/outputs of the module in DrummerNet. $C*$ indicates the number of channels but unspecified.}
\label{table:structure}
\end{table}

\begin{figure}[t]
\centering
 \includegraphics[width=\columnwidth]{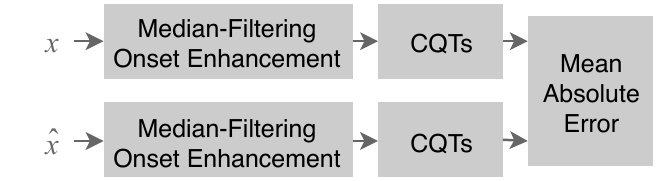}
 \caption{The block diagrams of loss calculation}
 \label{fig:onset_loss}
\end{figure}

\begin{figure}[t]
\centering
 \includegraphics[width=\columnwidth]{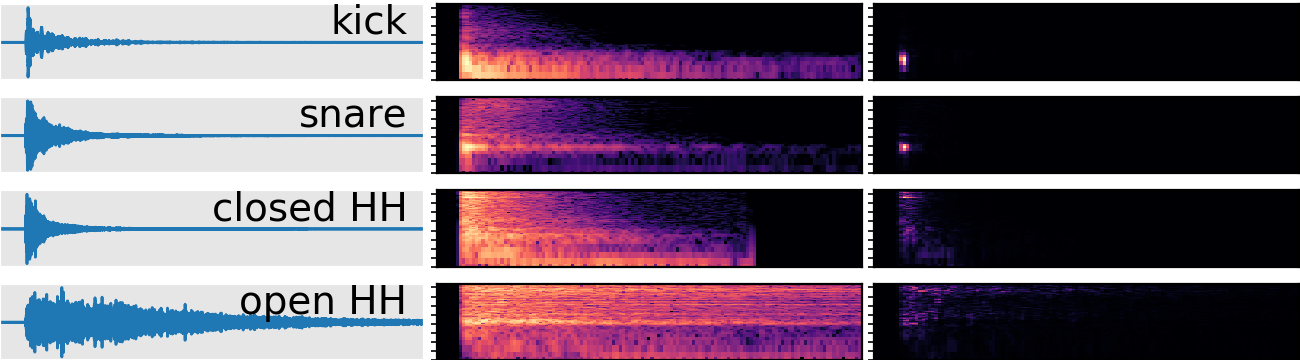}
 \caption{The effect of drum extraction for kick, snare, close hi-hat, and open hi-hat, from top to bottom. Columns are from left to right: original waveform, original spectrum, and onset-enhanced spectrum}
 \label{fig:onset_fx}
\end{figure}

\largesqueeze
\subsection{Learning}

Unable to directly compute the transcription loss during unsupervised learning, we carefully designed a loss function at the audio level, $L_x(x, \hat{x})$, as minimizing it would also minimize the transcription loss, $L_y(y, \hat{y})$. 
To do so, $L_x$ should be able to differentiate the drum components -- kick drum (KD), snare drum (SD), and hi-hat (HH) -- while being invariant to the varying drum kits.
Perceptually, there are clear differences between KD, SD, and HH. Although both impulsive, KD is in the low-frequency band while SD is in the mid-frequency band. SD is also relatively tonal and has a longer envelope. HH is more complicated to describe due to its variation from its playing technique. For example, closed and pedalled-HH's are in the high-frequency band, impulsive, and with relatively low energy, while open-HH's are similar except louder with a longer noisy envelope.

We thus define and use \textit{onset spectrum similarity}, which is designed to represent the similarity based on the onset part of sounds in the spectrum domain. As illustrated in \ref{fig:onset_loss}, it is measured by  \romannumeral 1) applying median-filtering based drum extraction \cite{fitzgerald2010harmonic} which enhances onsets (with a FFT size of 1024 and median filter length of 31 on both axes), \romannumeral 2) converting to multi-resolution CQTs (constant-Q transform) for both $x$ and $\hat{x}$, and then \romannumeral 3) calculating the mean absolute difference between them.

Among many spectral magnitude representations, we use (log-magnitude) CQT since the logarithmic frequency scale is known to match well to human auditory perception \cite{moore2012introduction}. We followed the implementation of Pseudo-CQT\footnote{
\url{http://librosa.github.io/librosa/}
} 
which multiplies linear-to-octave filterbanks to an STFT. As a result, the CQT covered nearly 8-octave bands from 32.07 Hz (\texttt{C1}) to 8 kHz (the Nyquist frequency of our experiment) with a 12-band/octave resolution. This implementation is differentiable.

Figure~\ref{fig:onset_fx} shows the effect of onset enhancement. It preserves the characteristics of the drum components in the transient part while removing the after-onset components. This process makes $L_x$ and $L_y$ more similar, as the non-transient parts vary more among drum kits due to their random and noisy nature. In a preliminary experiment, for example, the network tried to reconstruct all the non-transient components of SD using tom-toms and HHs, resulting in non-sparse and severe false-positive detection of onsets.

\smallsqueeze
\section{Experiments and Analysis} 
\label{sec:experiment}
\smallsqueeze
\subsection{Setup}

For the training of DrummerNet, we used an in-house dataset of drum stems that are crawled from many websites. The dataset consisted of 3,940~unique tracks averaging 225 seconds each for a total of 249~hours. Since the dataset was crawled from various websites, some details, such as the distribution of drum components, are hard to identify. The tracks were mostly popular western rock/pop music. Alternatives to this in-house dataset can be found in \cite{cartwright2018increasing} (3,758~drum sample recordings ($\times$8 second = over 8~hours) or 60,000~synthesized drum loops ($\times$8~second = over 133~hours)) and \cite{vogl2018towards} (4,197~drum tracks (259~hours)). We opted for the in-house dataset because it provided more diversity as it was not synthesized.

Each audio file was resampled to 16~kHz and downmixed to mono. The training batch size was 16, and for each audio file, we randomly selected a 2-second segment. On average, there were 112.5 segments in a track, and therefore training with 443,250 (=3,940 $\times$ 112.5) items would be approximately one epoch. With a Nvidia Tesla P100 and a batch size of 32, it took about 9~hours to train a single epoch. We implemented DrummerNet using Pytorch 1.0 \cite{paszke2017automatic} and used  Librosa 0.6.3 \cite{brian_mcfee_2019_2564164} and Madmom 0.16 \cite{madmom} for audio processing and peak-picking.

We used a heuristic peak-picking method introduced in \cite{bock2012evaluating}. This method selects a peak $\hat{y}[n]$ at $n$ if it satisfies the three conditions in Eq. \eqref{eq:1},

\squeeze
\smallsqueeze
\begin{equation} \label{eq:1}
\begin{aligned} 
        \hat{y}[n] =& \text{max}(x[n-w_m],...,x[n+w_m]) \\
    \hat{y}[n] \geq& \text{average}(x[n-w_a],...,x[n+w_a]) + \delta \\
    n >& n_{lp} + w_w, 
\end{aligned}
\end{equation}

where the max window $w_m$=50~ms, average window $w_a$=100~ms, threshold $\delta$=0.2, waiting window $w_w$=50~ms, and $n_{lp}$ is the last detected peak. We mainly use F1~score along with Precision and Recall using mir\_eval \cite{Raffel14mir_eval}. The tolerance window is 50~ms.

After training, we test the system on three public datasets: IDMT-SMT-Drums (SMT, 104 drum tracks, total 130 minutes  \cite{dittmar2014real}), Medley-DB Drums (MDB, 23 tracks, total 20 minutes \cite{southall2017mdb}), and ENST-drums (ENST, 61 minutes \cite{gillet2006enst}, drum-only tracks known as `wet-mix' of `minus-one' subset). According to \cite{wu2018review}, a task is DTD\footnote{DTD: drum transcription of drum-only recordings} if tracks are drum-only, more precisely KD/SD/HH-only, and the system annotates KD/SD/HH events. This is the case for the SMT dataset. A task with the system annotating KD/SD/HH but with drum tracks consisting of more than those three components, e.g., tom-toms and cymbals, is named DTP\footnote{DTP: drum transcription in the presence of percussion} in \cite{wu2018review}. Following this convention, we evaluate DTD with SMT (Section \ref{sec:relative_performance}), and DTP with MDB/ENST. We did not fine-tune for any dataset in any experiment and used the whole datasets for evaluation only.

\smallsqueeze
\subsection{Trend of Performance over Training}
\begin{figure}[t]
\centering
 \includegraphics[width=\columnwidth]{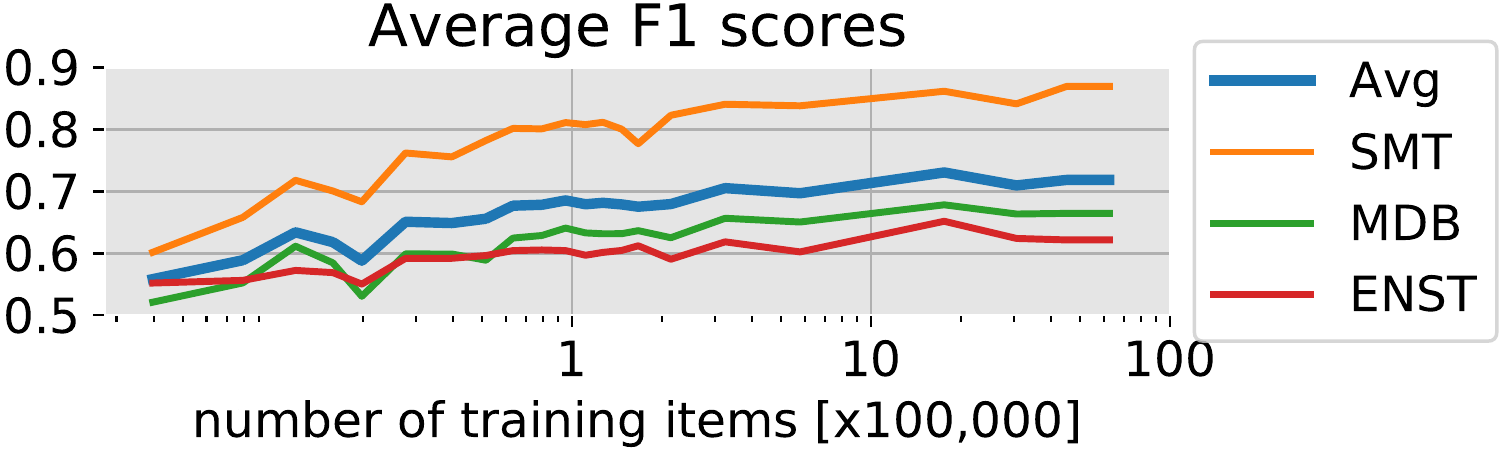}
 \caption{The F1 scores of DrummerNet over training items on each dataset (SMT, MDB, ENST), averaged over KD, SD, and HH. AVG indicates the overall average F1 scores of three datasets.}
 \label{fig:trend}
\end{figure}
\squeeze

We did not employ a stopping strategy but trained the network for $6 \times 10^6$ items (about 13 epochs). As illustrated in Figure \ref{fig:trend}, the overall performance gradually increases as the training proceeds and approaches converging towards the end of training. This indicates that the proposed loss function is a good proxy of transcription loss. After the initial phase of training, the performance differences among datasets remain consistent, probably due to the different characteristics of drum tracks in each dataset, as will be discussed in Section \ref{subsec:qual_a}.

\smallsqueeze
\subsection{Relative Performance against Baselines}\label{sec:relative_performance}
In this experiment, we trained our system on the in-house training set without any annotation and evaluated it on a separate test set (also known as `eval-cross (trained on DTP)', \cite{wu2018review}), which is a stronger condition than a usual train/test split scenario in supervised learning (`eval-subset', \cite{wu2018review}). This setup allows us to measure the generalization capabilities across the datasets. Specifically, our experiment is equivalent to DTD, `eval-cross (trained on DTP)' experiment in \cite{wu2018review}.\footnote{
Numbers are omitted in the paper but are available online: \url{https://www.audiolabs-erlangen.de/resources/MIR/2017-DrumTranscription-Survey}.
}, which is only available on SMT. Therefore, only the performances on SMT are compared in this Section. 
Overall, the performance of DrummerNet is favorable to that of recent drum transcription systems. With an average F1 score of 0.869 on SMT, the proposed unsupervised DrummerNet outperformed 9 out of 10 systems. The nine systems include ones with deep neural networks and supervised approach (ReLUts, RNN, lstmpB, tanhB, and GRUts \cite{vogl2016recurrent, southall2017automatic, southall2016automatic, vogl2017drum}),
as well as ones with NMF and unsupervised approach (AM1, AM2, PFNMF, and SANMF \cite{dittmar2014real, wu2015drum}).
It did not outperform NMFD \cite{lindsay2012drumkit}, a system based on the convolutive NMF. 

The comparison between DrummerNet and the NMF/unsupervised learning-based systems \cite{dittmar2014real, wu2015drum} implies that the proposed deep neural network structure effectively learns relevant representations. Furthermore, DrummerNet allows constant-time inference, unlike NMF and other factorization-based approaches which require iterative optimization in the test time.

\begin{figure}[t]
\centering
\includegraphics[width=\columnwidth]{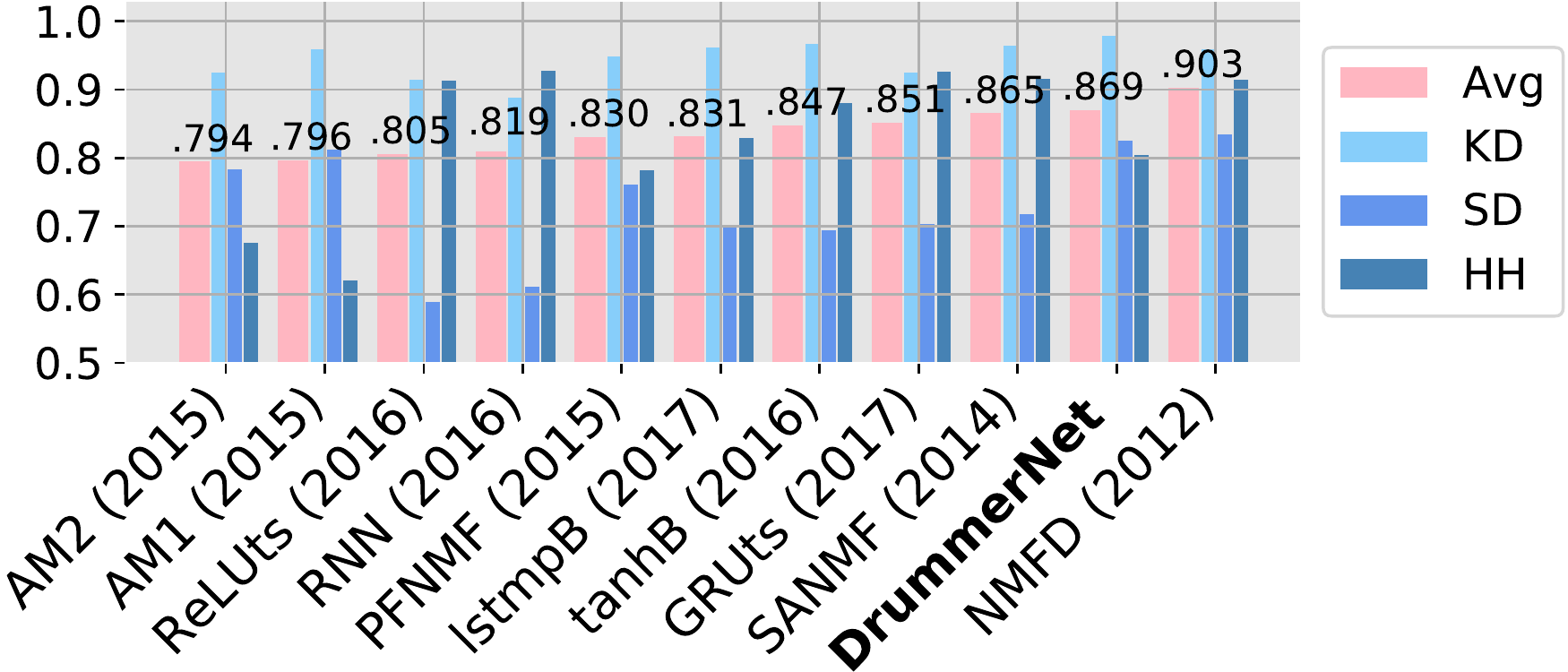}
 \caption{The F1 scores of DrummerNet and other systems on SMT with `eval-cross' setup, sorted by the ascending order of the overall average. The system names follow \cite{wu2018review}.}
 \label{fig:result_compare}
 \squeeze
\end{figure}

What is more interesting is its \textit{generalizability}. 
All the deep learning based systems\footnote{RNN, tanhB, ReLUts, lstmpB, GRUts - RNN-based systems} present deteriorating performance in the transfer learning scenario (eval-cross) compared to the dataset split scenario (eval-subset).\footnote{See Figure~10~(b) of \cite{wu2018review}. Note that most of the reported scores in papers also follow eval-subset setup.} However, less data-driven approaches\footnote{SANMF, NMF, PFNMF, AM1, AM2 - NMF-based systems} present similar or even increased performances in eval-cross. This implies that the distributions within datasets are fairly different and biased to certain types of drum tracks
and therefore, a transcription system trained with those datasets will be also biased accordingly. This limitation may be attributed to the small sizes of those datasets. Theoretically, supervised deep learning systems may generalize better if trained on a very large dataset, which lacks practicality due to the high annotation cost. In contrast, it is relatively easy to \textit{unbias} DrummerNet. One only needs to control the distribution of drum tracks by their style/genre/sounds without annotating every note.

%
%

\begin{figure}[t] 
\centering
    \begin{subfigure}{\columnwidth}
            \centering
            \includegraphics[width=\textwidth]{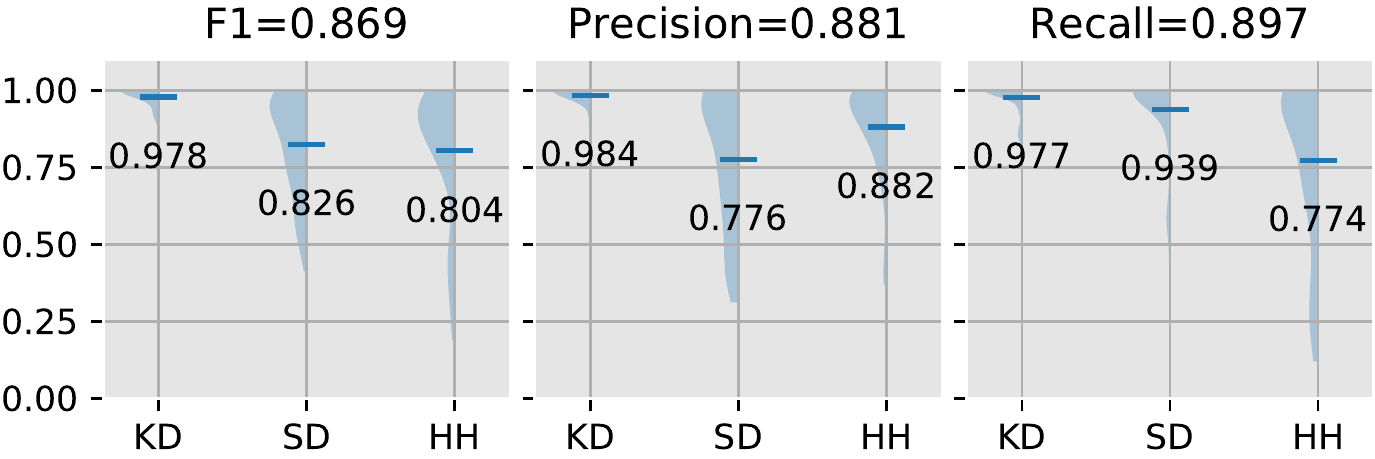}
    \end{subfigure}

    \begin{subfigure}{\columnwidth}
            \centering
            \includegraphics[width=\textwidth]{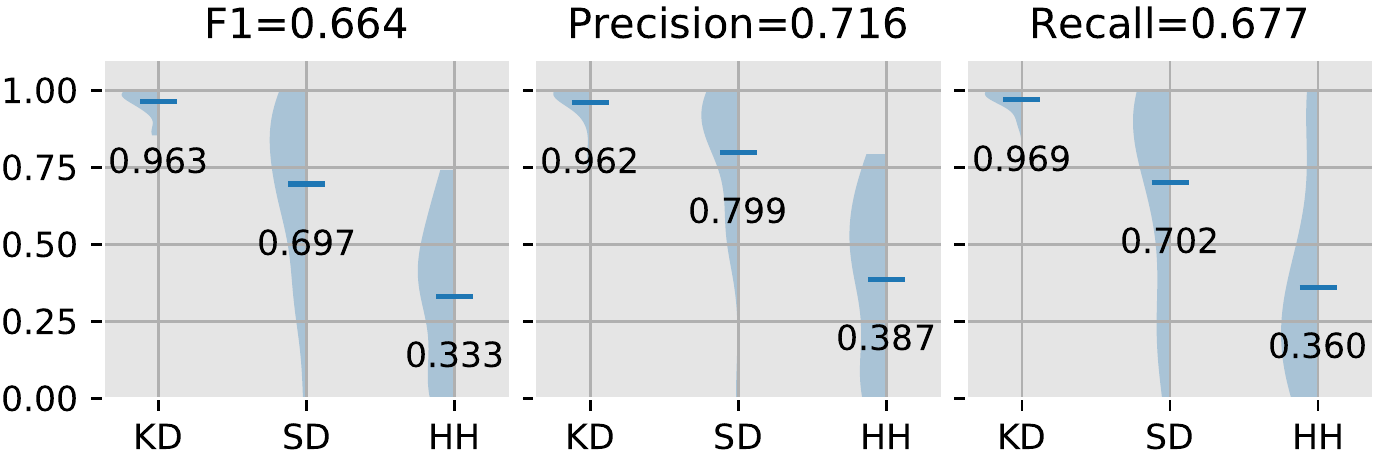}
    \end{subfigure}

    \begin{subfigure}{\columnwidth}
            \centering
            \includegraphics[width=\textwidth]{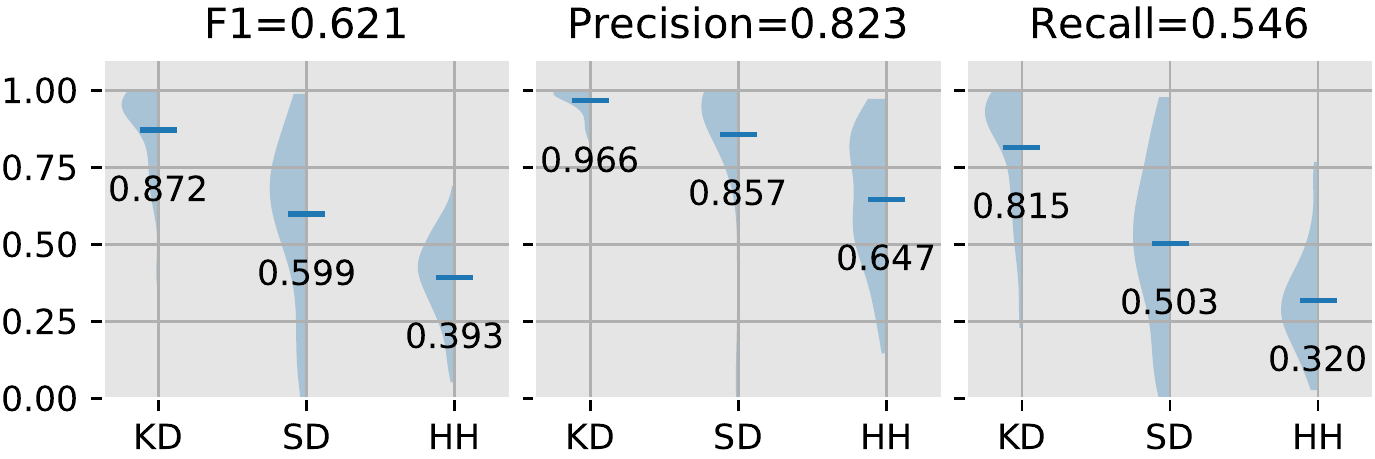}
    \end{subfigure}
    \caption{Evaluation of DrummerNet on SMT (top), MDB (middle), and ENST (bottom) datasets.}
    \label{figure:all_results}
\end{figure}

\begin{figure}[t]
\centering
 \includegraphics[width=\columnwidth]{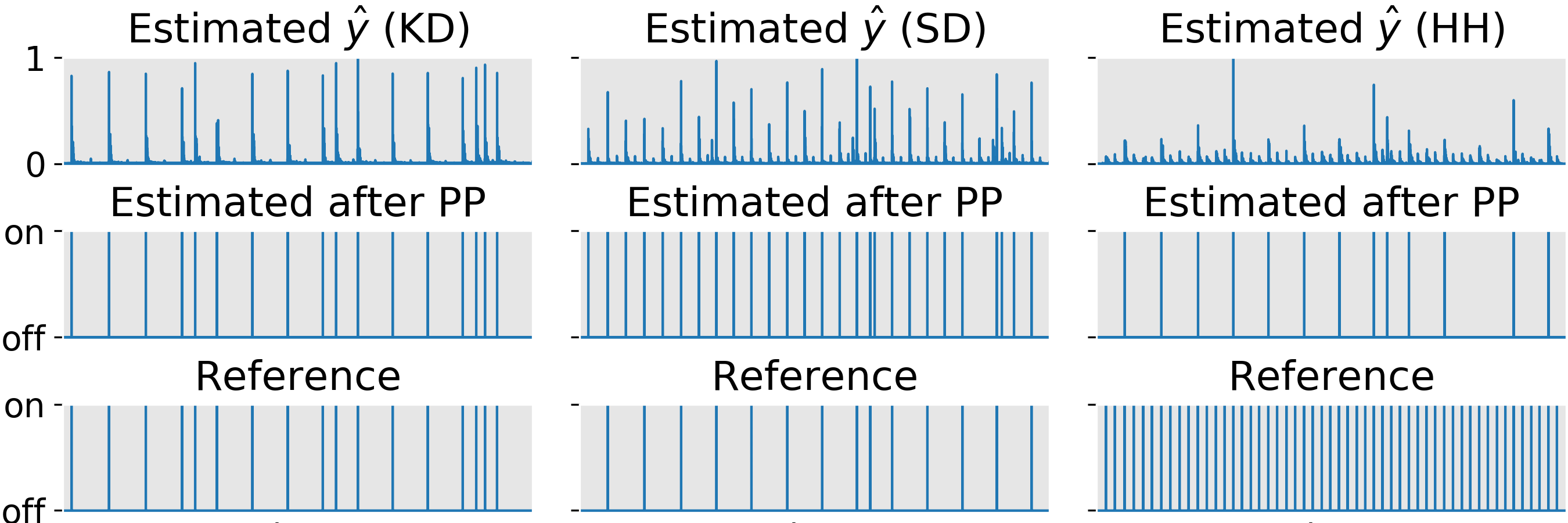}
 \caption{A transcription example of DrummerNet, `Real Drum 01-12' in SMT - the output of analysis module (top), after peak-picking (middle), and ground truth (bottom); KD, SD, HH (left to right).}
 \label{fig:my_problems}
\end{figure}

\largesqueeze
\subsection{Qualitative Analysis} 
\label{subsec:qual_a}

In this section, we will analyze the performance and the behavior of DrummerNet 
by components, datasets, and metrics, as illustrated in Figure~\ref{figure:all_results}.
Here, we notice two clear trends. First, across all of the three datasets and the metrics, detecting KD was the easiest, followed by SD and HH (except the precision on SMT). Second, SMT seems to be the easiest, followed by MDB and ENST. \textit{What could be the reasons?}

The first trend is strongly related the proposed loss function. KD has the least within-class variability while being the most distinguishable component (the largest mutual-class variability) due to its solitary frequency range. SD and HH share both the mid and high-frequency ranges and their sounds can vary significantly across drum kits -- i.e., larger within-class variability and smaller mutual-class variability. A common pattern, consequently, is the false positive of HH due to SD and vice versa. This is presented in Figure~\ref{fig:my_problems}, where SD has many false positives due to HH.

The second trend is caused by the mixed use of the probability and the onset velocity in the DrummerNet. Although transcription $\hat{y}$ \textit{is} the estimated amplitude of drum components, the peak-picking method treats $\hat{y}$ as if it was a probability. This discrepancy becomes problematic when the velocities of drum events in a track vary drastically as in the case of MDB and ENST. A failure case is demonstrated in Figure~\ref{fig:my_problems}, where the HH with strong accents on several occasion caused DrummerNet to miss many of the other HH peaks.

\begin{figure}[t]
\centering
 \includegraphics[width=0.94\columnwidth]{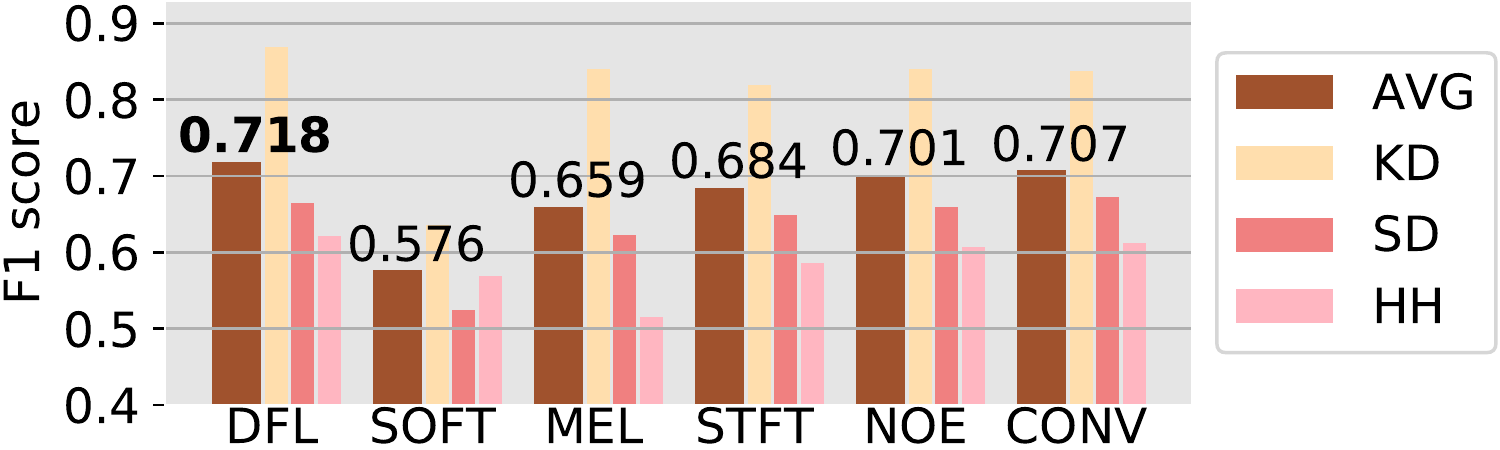}
 \caption{The ablation study results, F1 scores averaged over three datasets per component (KD, SD, HH) and their overall average (AVG). The label indicates as follow: DFL~(default DrummerNet as introduced), SOFT~(two Softmax layers instead of Sparsemax), MEL~(use 128-band melspectrogram instead of CQTs), STFT~(use 1024-point STFT instead of CQTs), NOE~(not onset enhancement in loss), CONV~(3-layer convolutional layers instead of recurrent layers).}
 \label{fig:ablation}
\end{figure}

\largesqueeze
\subsection{Ablation Study}

We conducted an ablation study where the performance of DrummerNet is compared with that of its variants. Figure~\ref{fig:ablation} shows the reported F1 scores averaged over datasets and components. Please refer to the caption in Figure~\ref{fig:ablation} for the definitions of the system names.

\begin{figure}[t]
\centering
 \includegraphics[width=\columnwidth]{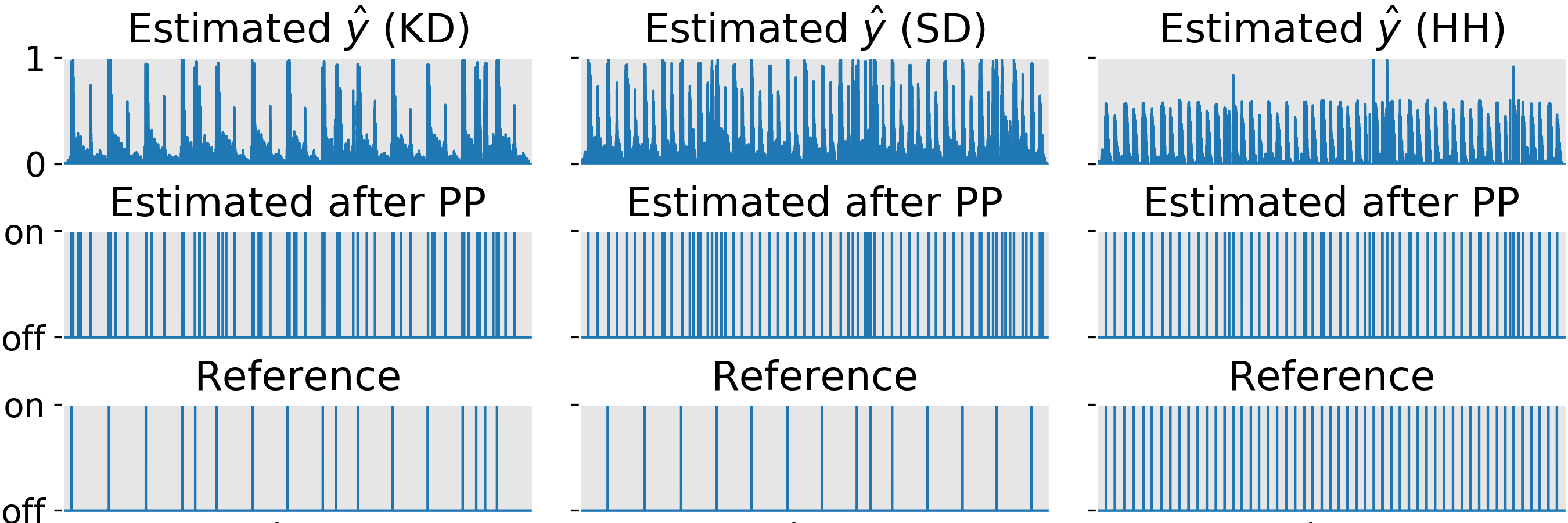}
 \caption{A transcription example of SOFT (DrummerNet with Softmax) , `Real Drum 01-12' in SMT - the output of analysis module (top), after peak-picking (middle), and ground truth (bottom); KD, SD, HH (left to right).}
 \label{fig:soft_example}
\end{figure}

\largesqueeze
\paragraph{Sparsemax (DFL vs. SOFT)}

Among all the variants in this experiment, we observe the most dramatic change in the performance when we replaced Sparsemax with Softmax (SOFT), mostly in a negative way. In SOFT, the two Softmax layers were applied in sequence instead of in-parallel and multiplied, which we tested, but the training was unstable. The transcription $\hat{y}$ of SOFT tends to be much noisier with many false positives, as presented in Figure \ref{fig:soft_example}. We conclude that the sparsity induced by Sparsemax is a crucial factor behind the success of the proposed unsupervised transcription.

Figure~\ref{fig:soft_example} provides a good example of the performance degradation pattern for each component. As in Figure~\ref{fig:ablation}, although the scores of all the three components decrease in SOFT, the degradation is not as critical for HH as in the case of KD/SD. This observation reflects the underlying properties of the different components. KD and SD are sparser than HH, and thus may benefit more from the introduction of Sparsemax. 

\largesqueeze
\paragraph{CQT (DFL vs. MEL vs. STFT)}

Replacing CQTs with either melspectrograms (MEL) or short-time Fourier transform magnitudes (STFT) results in decreased performance. Unlike CQTs, where different numbers of FFT are used for each octave range, melspectrograms are computed based on single-resolution STFT. This implies that  DrummerNet benefits from CQTs which consider multiple temporal and spectral resolutions.

Comparing MEL and STFT, the melfrequency compression helps with the better detection of KD but not SD nor HH. This is explained by the different frequency band weighting of STFT and melspectrogram. Since melfrequency is linear below 1~kHz and logarithmic above 1~kHz \cite{slaney1998auditory}, melspectrogram allocates relatively more bins below 1~kHz. This means that the loss function in MEL is biased towards the low-frequency range, resulting in training that favors KD over the others.

\largesqueeze
\paragraph{Onset Enhancement (DFL vs. NOE)}

The onset enhancement is shown to be boosting the performance, but not significantly (0.017). In the learning curve, we observe that removing the onset enhancement from the loss function results in a large performance degradation during the initial phase of training. This is mainly due to false-positives in the non-transient part.

\largesqueeze
\paragraph{Recurrent layers (DFL vs. CONV)}

Overall, replacing three recurrent layers with three convolutional layers does not make significant differences (0.011). This may means i) a long-term relationship may not provide additional information, probably because the transcription largely depends on local information, and ii) the mutual conditioning in the last recurrent layer is not effective in our experiment. In an informal analysis, we observed that with recurrent layers, $\hat{y}$ still has some local temporal correlation, e.g., the activations are smeared over time, probably because that is better to reconstruct the input audio. 

\section{Conclusion} \label{sec:conclusion}
We introduced DrummerNet, a deep neural network that is trained to transcribe drum tracks without a labeled dataset. In the experiment, DrummerNet achieved strong performance compared to existing systems trained with supervised learning, showing its generalizability towards a real-world drum transcription scenario. Our ablation study showed that Sparsemax and CQT played a crucial role in the successful training of DrummerNet.

The experiment also revealed room for further improvements. Considering the discreteness of the musical notes, a reinforcement learning approach may be more suitable \cite{southall2018player}, making the prediction more sparse and replacing the peak-picking with trainable action. 
The onset-enhancement on audio similarity is a function carefully-chosen in order to approximate $L_y$ when $x$ and $\hat{x}$ are given. Unfortunately, the approximation is limited because the exact drum sounds in $x$ are not given, and therefore a perfect reconstruct of (onsets of) the input audio ($L_x(x, \hat{x})=0$) does not lead to a perfect transcription ($L_y(y, \hat{y})=0$). An alternative way would be measuring a similarity on a (perceptual) representation domain instead of the audio, for example, by learning a loss using forward-backward consistency (also known as a cyclic loss \cite{kalal2010forward}) or known audio features.
Lastly, the current synthesizer module is limited to drums as it does not handle the duration of notes. A trainable synthesizer can be used to expand DrummerNet to other instruments \cite{engel2017neural, blaauw2017neural}, eventually leading to an unsupervised universal transcription system combined with instrument recognition.

\section{Acknowledgement}
We thank Tristan Jehan and Sebastian Ewert for their valuable comments and discussions. We would also like to express our sincere gratitude to Chih-Wei Wu for sharing his insight with us. 

\bibliography{ISMIRtemplate}
\end{document}